\documentclass{article}
\usepackage{ijcai24}
\usepackage{placeins}
\usepackage{times}
\usepackage{soul}
\usepackage{url}
\usepackage[hidelinks]{hyperref}
\usepackage[utf8]{inputenc}
\usepackage[small]{caption}
\usepackage{graphicx}
\usepackage{amsmath}
\usepackage{amsthm}
\usepackage{amsfonts}
\usepackage{booktabs}
\usepackage{algorithm}
\usepackage{algorithmic}
\usepackage{nicefrac}
\usepackage[switch]{lineno}
\usepackage{enumitem}
\usepackage[capitalize]{cleveref}

\usepackage{subcaption}
\usepackage{quoting}
\usepackage{epigraph}
\usepackage{todonotes}
\usepackage{comment}
\usepackage{appendix}

\usepackage{tikz}

\newtheorem{observation}{Observation}

\newtheorem{corollary}{Corollary}

\newtheorem{theorem}{Theorem}

\newtheorem{remark}{Remark}
\newtheorem{definition}{Definition}
\newtheorem{example}{Example}

\newcommand{\argmax}{\mathop{\mathrm{argmax}}}

\newcommand{\calF}{\mathcal{F}}

\newcommand{\brackets}[1]{\left(#1\right)}

\newcommand{\set}[1]{\left\{#1\right\}}

\newcommand{\Sum}[2]{\overset{#2}{\underset{#1}{\sum}}}

\title{Optimizing Viscous Democracy}

\author{
Ben Armstrong$^1$
\and
Shiri Alouf-Heffetz$^2$\and
Nimrod Talmon$^{2}$
\affiliations
$^1$University of Waterloo\\
$^2$Ben-Gurion University
\emails
ben.armstrong@uwaterloo.ca,
shirihe@post.bgu.ac.il,
talmonn@bgu.ac.il
}

\begin{document}

\maketitle

\begin{abstract}

\emph{Viscous democracy} is a generalization of liquid democracy, a social choice framework in which voters may transitively delegate their votes. In viscous democracy, a "viscosity" factor decreases the weight of a delegation the further it travels, reducing the chance of excessive weight flowing between ideologically misaligned voters.
We demonstrate that viscous democracy often significantly improves the quality of group decision-making over liquid democracy. We first show that finding optimal delegations within a viscous setting is NP-hard. However, simulations allow us to explore the practical effects of viscosity. Across social network structures, competence distributions, and delegation mechanisms we find high viscosity reduces the chance of ``super-voters'' attaining large amounts of weight and increases the number of voters that are able to affect the outcome of elections. This, in turn, improves group accuracy as a whole. As a result, we argue that viscosity should be considered a core component of liquid democracy.

\end{abstract}

\section{Introduction}

Liquid democracy~\cite{blum2016liquid}, an evolving concept in democratic theory, seeks to address the limitations of traditional direct and representative democracy by enabling flexible delegation of voting power. To do so, in liquid democracy, individuals choose between casting their votes directly or delegating their voting power to another voter of their choice. In contrast to simple proxy voting~\cite{green2015direct}, delegations are transitive and a delegation may ``flow'' through many voters. By merging the strengths of direct democracy and representative decision-making, liquid democracy enhances epistemic decision-making by incorporating diverse perspectives and informed opinions.

We study a generalization of liquid democracy -- viscous democracy~\cite{boldi2011viscous} -- which effectively limits the maximum weight voters can attain. In this variant, vote weights decrease exponentially the further a delegation travels. This ``viscosity'' in the computation of the vote weights has its roots in random walks and the PageRank algorithm~\cite{berkhin2005survey}. Viscous democracy offers a compromise between the advantages of delegation and direct voting, potentially curbing excessive concentration of power while maintaining the flexibility core to liquid democracy.

Here, we are interested in quantifying the potential benefit of adding viscosity to liquid democracy. A pivotal factor in this assessment is the dampening factor or ``viscosity'' -- denoted by~$\alpha$ throughout the paper -- which governs the rate of weight decay as delegation chains lengthen.
As we show, the influence of $\alpha$ extends beyond decision quality, affecting the overall structure of those delegation graphs that achieve a high quality of decision making. We are interested in identifying suitable values of~$\alpha$ generally and for specific settings.
Specifically, we seek to unravel the relationship between the value of $\alpha$ and the quality of decision making, as well as exploring how $\alpha$ shapes the topology of optimal delegations.

In this paper we adopt a dual approach, employing analytical and theoretical methodologies, as well as computer-based simulations. Theoretical analyses provide insights into the fundamental dynamics and properties of viscous democracy; simulations, on the other hand, offer a tangible platform for observing the empirical behavior of viscous democracy under varied conditions, facilitating a more realistic assessment of its outcomes and interactions. The combination of these approaches enriches our understanding of the complex interactions within the system of viscous democracy.

We view our work as fundamental to the better understanding of viscous democracy; as such, it paves the way to tools that can optimally parameterize viscous democracy in different settings, setting the dampening parameter $\alpha$ based on the specific properties of the setting at hand. As we show, the improvement to the quality of decision making from viscosity is frequently significant.

\begin{remark}
The introduction of viscosity to liquid democracy is a deviation from the principle of ``one person, one vote'' which, therefore, may render it inappropriate for democratic contexts in which equality in voting is essential~\cite{shapiro2018point}. We show, however, that it does prove advantageous in other applications; in particular, in those that prioritize prediction accuracy (e.g., a group prediction regarding the price of some stock), group recommendation (the application discussed alongside the introduction of viscous democracy~\cite{boldi2011viscous}), and other, general collaborative settings (e.g., a collaborative venture capital such as The DAO~\cite{hassan2021decentralized}). In these scenarios -- due to varying levels of competence in the population around different topics of discussion -- the nuanced decision-making facilitated by viscosity enhances collective intelligence, allowing for better decision making.
\end{remark}

\section{Related Work}

By now, there is a relatively large corpus of literature on liquid democracy. 
This includes works that study liquid democracy from a political science perspective~\cite{blum2016liquid,paulin2020overview}; from an algorithmic point of view~\cite{kahng2021liquid,dey2021parameterized,caragiannis2019contribution}; from a game-theoretic point of view~\cite{zhang2021power,bloembergen2019rational,escoffier2019convergence}; and from an agenda aiming to increase the expressive power that is granted to voters~\cite{brill2018pairwise,jain2022preserving}.

This work concentrates on the epistemic approach to social choice; we assume some ground truth of the decision to be made and evaluate the accuracy of the voting system at revealing ground truth using a common model for epistemic liquid democracy \cite{alouf2022should,kahng2021liquid,caragiannis2019contribution,halpern2021defense}.

Viscous democracy is relatively under-studied within liquid democracy. The original paper on the topic \cite{boldi2011viscous} establishes some intuition and a relation to the PageRank algorithm \cite{berkhin2005survey}; but there is no analysis regarding the effect of one specific value of the dampening factor compared to another, nor is there analysis discussing how to choose the factor. Viscous democracy was also applied as a setting for group recommendation system~\cite{boldi2015liquid}.

\section{Model}

Our formal model describes how voters are embedded in an underlying social network, how they may delegate, and how group decisions are ultimately made.

\subsection{Delegation Graphs}

We adopt a basic epistemic social choice setting with $n$ voters, $V = \{v_1, \ldots, v_n\}$, and two alternatives $A = \{a^+, a^-\}$. We assume that $a^+$ is an objectively correct outcome that all voters collectively aim to elect. Each voter $v_i \in V$ has a competency level $q_i \in Q$ in the range $[0, 1]$ that corresponds to the probability that $v_i$ would vote ``correctly'' (i.e., would vote for $a^+$). Voters are connected via an underlying social network $G = (V, E)$ with a set of undirected edges $E$ representing the connections between voters (where delegations can occur as direct arcs only along these connections).

Our model generalizes liquid democracy -- a system enabling transitive delegations. Each voter $v_i$ is able to perform one of two actions: They may vote directly, as above, supporting $a^+$ with probability $q_i$, or they may delegate their vote to a voter in their set of neighbours, $N_G(i) = \{j \in V | (i, j) \in E\}$. If a voter votes directly, then we refer to them as a \textit{guru} and model them as delegating to themselves.

A delegation function $d: V \rightarrow V$ outputs the delegation of each voter. $d(v_i) = v_j$ indicates that $v_i$ delegates to $v_j$. Delegation applies transitively so that a delegation might ``travel'' several hops before reaching a guru. $d^*(v_i)$ refers to the repeated application of $d(v_i)$ until a self-delegation (i.e. a guru) is reached. The guru of $v_i$ is $d^*(v_i)$ and denote the set of all gurus as $\mathcal{G}(V)$. 
We disallow any delegation from $v_i$ to $v_j$ that would result in a cycle.
This is done to prevent loss of votes within the voting process. In our experiments we choose a fixed proportion of voters to delegate while the remainder vote directly. 

Note that a delegation function induces a directed subgraph of~$G$ containing (directed versions of the) edges in~$E$ only when a delegation flows between the two nodes of an edge. We call this a \textit{delegation graph} of $G$, denoted $D = (V, \{(i, j) \in E | d(i) = j \})$. A delegation graph can be conceived of as a forest of directed trees, where all edges in each tree flow towards the singular guru in that tree.

\subsection{Vote Weights}

In ``traditional'' liquid democracy, the weight of each guru is equal to the number of delegations it receives, either directly or indirectly. This paper builds upon a model introduced by \citeauthor{boldi2011viscous} referred to as \textit{viscous democracy} \cite{boldi2011viscous}. Each edge that a delegation travels reduces its weight by a constant dampening factor $\alpha \in [0, 1]$, also referred to as the viscosity. Thus, the weight of each voter $v_i$ is,

\[
w_{i} = 
\begin{cases}
0 & \text{if } v_i \notin \mathcal{G}(V) \\
\sum_{p \in \operatorname{Path}(-, i)} \alpha^{|p|} & \text{otherwise }
\end{cases}
\]

where $\operatorname{Path}\brackets{-, i}$ denotes all paths of delegations that reach $v_i$. 
Note the following:
\begin{itemize}

\item
When $\alpha = 1$, viscous democracy reduces to the standard liquid democracy setting. 

\item
At lower values of $\alpha$, guru weights change based upon the structure of the delegation graph. Weights of voters that are further away from their guru decrease more than the weights of voters close to the guru.

\item  As there are no cycles, there is at most one path between any two voters.

\end{itemize}

\subsection{Group Accuracy}

As there are only two alternatives, we use weighted plurality voting (following May's theorem~\cite{may1952set}) to determine the winner of an election. Specifically, each guru selects one alternative, based upon their competence, and commits all of their weight to supporting that alternative. The alternative receiving the most total weight is the winner. The probability that~$a^+$ will be selected as the winner is referred to as \textit{group accuracy}, or, simply \textit{accuracy}.

Accuracy is a function of the delegations and the voter competencies but, crucially, it is also affected by $\alpha$. We denote the accuracy of given viscosity, competencies, and delegations as $Acc(\alpha, Q, D)$. We refer to the value of $\alpha$ that maximizes group accuracy as $\alpha^*$. The existence of $0 < \alpha^* < 1$ and the dynamics of $\alpha^*$ are a major focus of this paper.

We can now formally specify an \textit{election} by either of two forms (we use these interchangeably, as the first form implies the second): 
\begin{enumerate}

\item
$\mathbb{E} = (Q, G, d)$:
Here, an election is a combination of voter competencies, an underlying social network, and a delegation function which may be used to determine a delegation graph $D$; or

\item
$\mathbb{E} = (Q, D)$:
Here, an election consists of voter competencies and an already existing delegation graph.

\end{enumerate}

\begin{figure}
    \centering
    \begin{tikzpicture}[
    roundnode/.style={circle, draw=blue!60, fill=blue!5, very thick, minimum size=4mm},
    squarednode/.style={rectangle, draw=red!60, fill=red!5, very thick, minimum size=4mm},
    ]
        \node[roundnode] (a1) {};
        \node[roundnode] (a2)[right= of a1] {} ;
        \node[roundnode] (a3)[right= of a2] {} ;
        \node[roundnode] (a4)[right= of a3] {} ;
        \node[squarednode] (A) [right= of a4] {A} ;
        \node[roundnode](b1)[below=0.5cm of a1]{};
        \node[squarednode] [right= of b1] (B){B};
        \node[roundnode][below=0.5cm of a4](c1){};
        \node[squarednode][right= of c1](C){C};
    
        \draw[->] (a1) -- (a2) -- (a3) -- (a4) -- (A);
        \draw[->] (b1) -- (B);
        \draw[->] (c1) -- (C);
        
    \end{tikzpicture}
    \caption{A Stars and Chains delegation graph. The value of $\alpha^*$ depends greatly upon the competence values assigned to each of gurus $A$, $B$, and $C$.}
    \label{fig:stars_and_chains_tikz_example}
\end{figure}
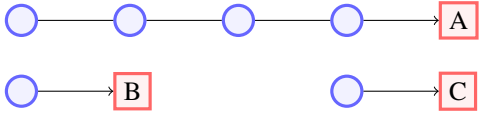

\section{Families of Delegation Graphs}

Throughout this paper we explore several different families of delegation graphs of two varieties: We first describe delegation graphs that are induced by the application of a delegation function to voters on a social network. Second, we introduce two simple delegation graph models that are constructed directly, without a specific delegation function, to serve as illustrative examples.

\subsection{Social Networks}

Our experiments explore two types of underlying social networks $G$ that voters exist upon: both empirical and artificially-generated networks. The empirical networks we consider are collected from a variety of sources and detailed in \cref{appendix:real_world_networks}.

To generate artificial networks we use two well-studied probabilistic models that are often used to replicate properties of real-world social networks~\cite{Kleinberg-networks-book}: Erdős–Rényi (ER) and Barabási–Albert (BA) networks. Both of these models are parameterized by the number of nodes and a parameter controlling their edge density ~\cite{erdos1960evolution,albert2002statistical}.
Unless stated otherwise, voters are placed on social networks uniformly at random.

\subsection{Delegation Functions}

Delegation functions use a social network and the competence of each voter to determine, for each delegating voter, to whom they delegate. We define delegation probability functions $p(v_i, v_j)$ as giving the probability that $v_i$ will delegate to $v_j$. A delegation function $d$ specifies a delegate for each voter based upon $p(v_i, v_j)$.
Our experiments consider several concrete delegation probability functions, defined formally in \cref{appendix:delegation_function_descriptions}:

\begin{itemize}
    \item \textbf{Max}: Each delegating voter delegates to their most competent neighbour.
    \item \textbf{Random Better}: Each delegating voter delegates to a neighbour more competent than themselves chosen with uniformly random probability.
    \item \textbf{Proportional Better}: Each delegating voter delegates to a neighbour more competent than themselves chosen with probability proportional to the competence of each potential guru.
    \item \textbf{Proportional Weighted}: Each delegating voter delegates to a neighbour more competent than themselves with probability proportional to the competence \textit{and} weight of each potential guru, such that gurus with higher weight are less likely to receive delegations.
\end{itemize}

\subsection{Delegation Graph Models}
\label{section:delegation_models}

We also explore two directly-constructed graph families which we use to demonstrate the possible effects of changing viscosity.
The first model is a natural adaptation of the model of preferential attachment while the second model generates very simple graphs that highlight some basic topological properties that affect the optimal viscosity.
\begin{itemize}

\item 
\textbf{Competence-Based Attachment (CBA)}:
This model creates a delegation graph by applying the idea of preferential attachment to voter competencies~\cite{battiston2020networks}. Given voter competencies and some set of gurus the remaining voters select a delegate based upon the relative competence of each potential delegate. For completeness, \autoref{alg:cba_algorithm} outlines the procedure in more detail.

\begin{algorithm}[tb]
\caption{Competence-Based Attachment (CBA)}
\label{alg:cba_algorithm}

\raggedright
\textbf{Input}: $V, Q, \mathcal{G}$

\begin{algorithmic}[1] 
\STATE $S \gets \mathcal{G}$
\STATE $E \gets \emptyset$
\STATE $D \gets (V, E)$
\WHILE {$S\neq V$}
    \STATE Select one voter $v_i\in V\setminus S$ uniformly at random.
    \STATE Select one voter $v_j \in S$ with probability proportional to its relative competence; $P(v_j) \propto \frac{q_j}{\Sum{v_k\in S}{}q_k}$
    
    \STATE \textbf{Attach:} $E \gets E\cup (v_i, v_j)$.
    \STATE \textbf{Update}: $S \gets S\cup \set{v_i}$.
\ENDWHILE
\end{algorithmic}
\end{algorithm}

\item
\textbf{Stars and Chains (SC)}:
A basic structure parameterized by $(s, s_\text{comp}, n_s, c, c_\text{comp}, n_c)$. This model constructs a delegation graph consisting of $s$ small \textit{star} components each with $n_s - 1$ delegators and 1 guru, and $c$ large \textit{chains} with $n_c - 1$ delegators. Gurus are thus located at the center of the stars and at one end of the chains. 
Gurus in star and chain components have competence $c_\text{comp}$ and $s_\text{comp}$, respectively. Each chain component should have more voters than are in all star components combined, that is $n_c > sn_s$ (typically we have considered settings where $n_c = sn_s + 1$ but that is not required). \autoref{fig:stars_and_chains_tikz_example} visualizes a simple SC delegation graph.

\end{itemize}

\section{Optimal Delegation Graphs}

Here we provide bounds on the complexity of finding delegations within a viscous setting that maximizes group accuracy.
Informally, given a social network and some value of $\alpha$, we are interested in the combinatorial problem of picking for each voter whether she will vote directly or whether she will delegate to some other voter, and, if so, to whom.


\begin{definition}
In the \textsc{Optimal Delegation Graph} (ODG) problem we are given a network of voters $G = (V, E)$ and the task is to find a subgraph $D = (V, E_D \subseteq E)$ that is a viable delegation graph of $G$ (i.e. there are no cycles in $D$) and which maximizes the accuracy. 
\end{definition}

Next we analyze the computational complexity of ODG.
First, note that, for the case of $\alpha$, we can solve ODG in polynomial time.

\begin{theorem}
  For $\alpha = 0$, ODG is in P.
\end{theorem}

\begin{proof}
Observe that, for $\alpha = 0$, setting a voter to delegate to some other voter effectively removes that voter, as no weight is transferred through delegation. Thus, to solve ODG we can iterate over $k \in [n]\cup \set{0}$, which is the number of voters to remove. In particular, for each $k \in [n]\cup \set{0}$, look for the $k$ voters with the lowest accuracy and remove them (i.e. have them make arbitrary delegations). Now, compute the accuracy (this can be done in polynomial time using dynamic programming~\cite{becker2021can}). Finally, select the $k$ for which the accuracy is the highest.

As all voters are equally weighted, it is always strictly better to include a more competent voter in place of a less competent voter and removing some number of least competent voters will solve ODG.
\end{proof}

For $\alpha = 1$, however, \citeauthor{caragiannis2019contribution} (\citeyear{caragiannis2019contribution}) have shown that ODG is NP-hard.
%

\begin{corollary}
  For $\alpha = 1$, ODG is NP-hard.
  Furthermore, for this case, it is also NP-hard to approximate ODG to within a factor of $\nicefrac{1}{16}$~\cite{caragiannis2019contribution}.
\end{corollary}

Below, we provide the first results on the complexity of ODG for $0 < \alpha < 1$.

\begin{theorem}\label{Thm: ODG is NP-Hard}
    For $\alpha = \frac{1}{m}, m\in \mathbb{N}$, ODG is NP-Hard.
\end{theorem}

\begin{figure}
\centering
\includegraphics[width=0.41\textwidth,keepaspectratio]{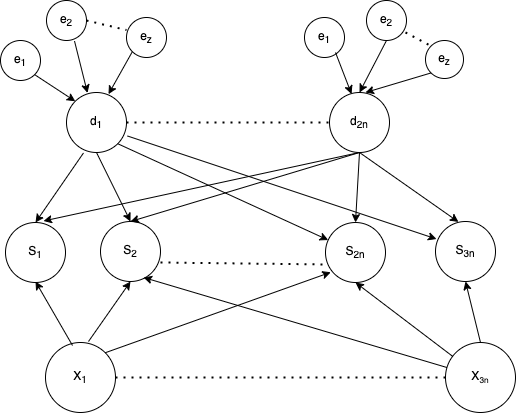}
\caption{Illustration of the reduction described in the proof of Theorem~\ref{Thm: ODG is NP-Hard}. There are 4 layers of nodes, the bottom layer is corresponding to elements from the set $X$, the second layer are the sets $S_1,..,S_{3n}$. The third layer is dummy voters $\set{d_i}$, and the fourth layer is an additional set of dummy voters connected to the third layer in order to adjust the weights.}
\label{fig:X3C_to_ODG}
\end{figure}

\begin{proof}
The proof follows a reduction from the NP-hard problem \textit{Restricted Exact Cover by 3-Sets (X3C)}~\cite{GONZALEZ1985293} to $ODG$.
An instance of X3C consists of a set $X = \set{x_1,..,x_{3n}}$, a family $\calF = \set{S_1,..,S_{3n}}$ where each set $S_i$ has 3 elements from $X$, and each element from $X$ appears in exactly 3 sets from $\calF$. The task is to decide whether a subfamily $\calF'\subset \calF$ such that $X = \underset{S_j \in \calF'}{\bigcup S_j} \text{and } S_j\cap S_i = \emptyset$ exists.

We will construct the following instance for the \textit{ODG} problem (see \autoref{fig:X3C_to_ODG}):
\begin{itemize}
\label{reduciton: X3C to ODG}
    \item For each set $S_1,.. ,S_{3n}$ we will have a voter $S_i$ with competence $q_{S_i} > 0.5$.
    \item For each element $x_i\in X$ we have 3 sets $S_{i_1}, S_{i_2}, S_{i_3}$ that $x_i$ appears in. We connect an outgoing edge to each $S_i$ from $x_i$. The competence of $x_i$ is set to $q_{x_i} = 0, \forall i\in{1,..,3n}$.
    \item Lastly, we have a set of dummy voters $d_1,..,d_{2n}$ with competence $q_{d} = 0$ that are connected to all the the \textit{set voters} $S_1,.., S_{3n}$ with outgoing edges. Each dummy voter has peripherally a set of dummy voters $e^{i}_1,..,e^{i}_z$ that are connected to it. Each has competence of $q_{e}=0$, designed to incentivize the dummy voters to delegate otherwise they contribute negatively to the outcome of an election. 
\end{itemize}

This completes the description of the reduction. Below we prove the two directions for correctness. 

$\implies$ Given a solution for the \textit{Restricted 3-Set Exact Cover} $\calF' = \set{S_{j_1}, S_{j_2}.., S_{j_n}}$, we shall delegate each $x_i$ to its corresponding $S_j$ it appears in $\calF'$. There is only one such set $S_{j_\ell}$ because $\calF'$ is an exact cover. Each of these $S_j$ in layer 2 (see figure \ref{fig:X3C_to_ODG}) voters now has a weight of $1+3\alpha$: weight from the node itself plus weight from three delegators, $x_{j_1}, x_{j_2}, x_{j_3}$, that are each one hop away (and thus have their weight reduced by the viscosity factor). For the remaining $2n$ sets in layer 2 we shall delegate one dummy voter from $d_1,..,d_{2n}$ to each set in layer 2 that is $S_r\in\calF\setminus \calF'$, the matching is arbitrarily. Since each set $S_r$ receives one dummy $d$ voter delegation and further delegations from $z$ peripheral $e$ voters that travel an extra hop. $S_r$ has a total weight of $1+\alpha + z\cdot \alpha^2$. Thus, we set $z = \frac{2}{\alpha}$ then all the $S_i$ voters in the central layer have the same weight, $w^{*}=1+3\alpha$. 
We know that this corresponds to the optimal weight distribution due to a direct application of Condorcet's Jury Theorem \cite{grofman1983thirteen}: Each $S_i$ has $q_{S_i} > 0.5$ so more active $S_i$ will monotonically increase group accuracy, and optimal weighted voting for when all $S_i$ have the same competence is equal weight distribution.

$\impliedby$ Next, we show that optimal delegation indeed leads to an exact cover. 
First we have to consider that the only voters that would be active are the $\set{S_i}$, since they are solely the ones with positive competence.  
They all have the same competence, that is larger then $0.5$, so an optimal outcome would take place when they all have equal weight, again (see \cite{grofman1983thirteen}).
Distributing the total weight equally would mean each \textit{set} voter $S_i$ would get one dummy voter $d_j$ with its associated voters $e^{i}_1,..,e^{i}_z$ or three delegations from  $x_1,..,x_{3n}$ voters. 
Since any other combination of voter delegations would mean some voters would receive excess weight and at least one voter would end up with less weight, thus causing a deviation from the optimal weight of $w^{*}=1+3\alpha$.
Note that we could achieve lower weight for $d_i$ if any of the peripheral $e^{i}_j$ did not delegate, but since their competence is zero they must not be active voters in the optimal delegation graph as that would reduce group accuracy.
Next we look at the sets $S_i$ that have incoming delegations from the first layer of $x_i$ and select those $S_i$'s to be our cover $\calF'$, we know that it satisfies all the condition to be a \textit{restricted exact 3 set cover} for $X$.  
\end{proof}

\begin{remark}
    Theorem \ref{Thm: ODG is NP-Hard} could be extended to any rational $\alpha=\frac{k}{m}$ by reducing from the $X\brackets{k+1}C$ problem (i.e., restricted exact cover with sets containing $k+1$ elements, where each element $x$ appear in exactly $k+1$ sets). The construction follows similar steps; we then set $z=m$, which causes the optimal weight to be $w^{*}=1+\brackets{k+1}\cdot\alpha$ for each set.
\end{remark}

\section{Some Pathological Cases Regarding \texorpdfstring{$\alpha^*$}{alpha^*}}

We now concentrate on whether viscosity provides actual benefits to accuracy over liquid democracy.
The standard model of liquid democracy used in most prior work only explores the situation where $\alpha = 1$. In contrast, recall that $\alpha = 0$ is a form of direct democracy where delegated votes become irrelevant as their weight is $0$ and all gurus have equal weight. Viscous democracy lies in the middle between these two settings with $0 < \alpha < 1$. In this section we show that there exist many parameterizations of delegation graphs for which a different one of direct, liquid, and viscous democracy are strictly beneficial.

Note that accuracy is a piecewise function with respect to~$\alpha$: i.e., accuracy changes only when $\alpha$ causes a change in the sets of gurus that make up a majority of weight. This means that, typically, $\alpha^*$ is not a unique value but rather a range of values. However, for simplicity we typically refer to it as a single value.




\subsection{Liquid Democracy: \texorpdfstring{$\alpha^* = 1$}{alpha = 1}}

We first show that there exist settings in which liquid democracy (i.e., where $\alpha^* = 1$) leads to optimal accuracy. Below we present an example where increasing $\alpha$ always weakly increases accuracy.

\begin{example}
\label{ex:3_voters-ld_best}

Consider the delegation graph in \autoref{fig:stars_and_chains_tikz_example} with 3 chains of voters. Let $q_A = 0.9$ and $q_B = q_C = 0.4$. When $\alpha = 1$, $w_A = 5$ and the outcome depends entirely on $A$'s vote and, thus, group accuracy is $0.9$. When $\alpha < 0.848$, $w_A < w_B + w_C$ and accuracy drops to approximately $0.772$.

\end{example}

\subsection{Direct Democracy: \texorpdfstring{$\alpha^* = 0$}{alpha^* = 0}}

We now show, for the topology of the delegation graph, a setting of competencies where lower values of $\alpha$ are weakly superior to higher values.

\begin{example}
\label{ex:3_voters-cd_best}
Consider again the SC delegation graph of \autoref{fig:stars_and_chains_tikz_example}. 
Let $q_A = 0.9$ and $q_B = q_C = 0.8$. As before, when $\alpha = 1$, group accuracy depends solely on $A$ and is $0.9$. In this case, however, decreasing $\alpha$ to the pivotal $\alpha = 0.848$ causes accuracy to increase due to the relative strength of voters $B$ and $C$. When $\alpha < 0.848$, accuracy increases to approximately $0.928$.
\end{example}

\subsection{Viscous Democracy: \texorpdfstring{$0 < \alpha^* < 1$}{0 < alpha^* < 1}}

In contrast to the previous examples we now present a simple example where accuracy is maximized when $0 < \alpha <1$.

\begin{example}
\label{ex:3_voters-vd_best}
Consider a SC delegation graph with $s = 6$, $s_\text{comp} = 0.8$, $n_s = 5$, $c = 3$, $c_\text{comp} = 0.5$, $n_c = 30$. That is, $6$ star components each with size $5$ and $3$ chain components each with size $30$.

When $0 \leq \alpha < 0.25$, the group accuracy is roughly $0.91$. However, the weight of star components increases more quickly than that of chain components as $\alpha$ increases. For $0.25 \leq \alpha \leq 0.5$, accuracy becomes approximately $0.94$. For higher values of $\alpha$, the larger chain components begin to dominate and accuracy decreases. Here, $\alpha^* \in [0.25, 0.5]$.

Figure~\ref{fig:non_extreme_alpha} showcases a family in which non-extreme alpha (i.e., $0 < \alpha^* < 1$) is optimal.

\begin{figure}[t]
    \centering
    \includegraphics[width=0.45\textwidth,keepaspectratio]{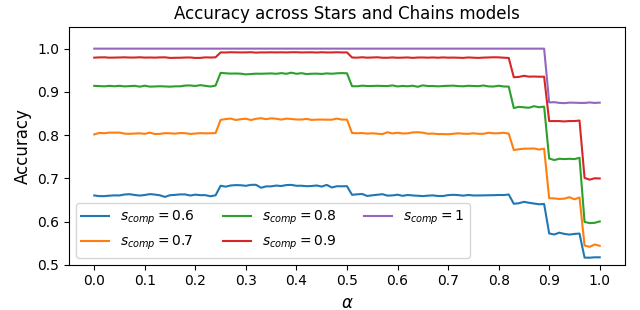}
    \caption{Accuracy in Stars and Chains delegation graphs as $\alpha$ varies from 0 to 1. Each series varies $s_\text{comp}$ and sets $s = 6$ $n_s = 5$, $c = 3$, $c_\text{comp} = 0.5$, $n_c = 30$ As $\alpha$ changes, the sets of gurus able to form a majority of weight shifts in a piecewise manner. Optimal $\alpha$ occurs in $[0.25, 0.5]$}
    \label{fig:non_extreme_alpha}
\end{figure}

\end{example}

\subsection{Intuition Regarding \texorpdfstring{$\alpha^*$}{alpha^*}}

Here we have demonstrated, for the first time, the existence of settings where $0 < \alpha^* < 1$; using the Stars and Chains delegation graph model. Intuitively, this graph structure leads to interesting dynamics as $\alpha$ changes due to the non-linearity in the corresponding changes in the weights of the gurus.

It is worthwhile to delve a little deeper into these graphs:
Observe that the weight of a guru $v_s$ in a star component increases linearly with $\alpha$: $w_s$ = $1 + n_s\alpha$. On the other hand, the weight of a guru $v_c$ in a chain component increases based on the polynomial $w_c = 1 + \alpha + \alpha^2 + \alpha^3 + ...$ with degree equal to the length of the chain. While this particular delegation structure is somewhat contrived, we believe similar structures may often occur in more realistic settings. In any setting where gurus receive the majority of their delegations at different distances from each other, there is the possibility that $\alpha < 1$ may be weakly, if not strongly, optimal.

\section{Experimental Analysis}
\label{sec:experiments}

We have shown by example that there exist delegation graphs with $\alpha^* < 1$. In this section we concentrate on the frequency with which $\alpha^*$ is below $1$. We examine first the CBA topology defined in \autoref{section:delegation_models} and then consider how often delegation mechanisms on other randomly-generated and real-world social networks benefit from viscosity.

In order to understand $\alpha^*$ under a wide range of settings, each of our experiments consider voters with competencies drawn from one of 3 distributions. These are often explored as 9 parameterizations of each distribution:
\begin{enumerate}[label=(\roman*)]
    \item \textit{Uniform} - Intervals of width $0.2$ from $U\brackets{0, 0.2}$ shifted up to $U\brackets{0.8, 1}$ in equal steps of $0.1$.
    \item \textit{Gaussian} - $\mu \in \{0.1, 0.2, ..., 0.9\}$, $\sigma = 0.05$\footnote{We use the SciPy implementation of the truncated normal distribution \cite{burkardt2014truncated}.}
    \item \textit{Exponential} - $\mu \in \{0.1, 0.2, ..., 0.9\}$\footnote{Since the exponential distribution does not provide an upper bound on sampled values whenever we sample a competency value greater than $1$ we map the value to $1$. This leads to a mean value slightly lower than the original distribution, i.e $\mu' = \frac{1}{\lambda} - \frac{e^{-\lambda}}{\lambda}$.}
\end{enumerate}

\subsection{Calculating Accuracy and \texorpdfstring{$\alpha^*$}{alpha^*}}\label{sec:optimal_alpha}

Before we can describe the results of our experiments, we discuss how we compute accuracy and $\alpha^*$.

\subsubsection{Monte Carlo Simulations for Group Accuracy}

As discussed by \cite{alouf2022should}, computing accuracy exactly for some liquid democracy settings is computationally impractical. Furthermore, the dynamic algorithm for computing accuracy described previously \cite{becker2021can} does not work with $\alpha < 1$. We use Monte Carlo simulation to estimate accuracy by running many elections with the same parameters. Each reported accuracy result is the proportion of 1000 elections in which $a^+$ was successfully elected, following the same procedure as \citeauthor{alouf2022should}.

\subsubsection{Estimating the Value of \texorpdfstring{$\alpha^*$}{alpha^*}}
Additional complexity that arises from the fact that \textit{group accuracy} is a piecewise function in terms of $\alpha$. Therefore the optimal $\alpha^*$ is actually an interval, in terms of notation we often refer to the upper bound of that interval. 
We are primarily interested in identifying settings where $\alpha^*$ is strictly below $1$ so we take a simple, cautious approach to estimation, described formally in \cref{appendix:estimating_optimal_alpha}. In short, we calculate accuracy several times at 21 evenly spaced values of $\alpha$ in $[0, 1]$ (chosen to test at every interval of 0.05). Given two values of $\alpha$: $\alpha_s$ and $\alpha_t$, the mean, and st. dev. of accuracy estimates at $\alpha_s$ and $\alpha_t$ we say that $\alpha_t$ is better than $\alpha_s$ if $\mu_t - \sigma_t > \mu_s + \sigma_s$. 

The best value emerging from this procedure is $\alpha^*$. If no value is strictly better than $\alpha = 1$ we say that $\alpha^* = 1$. As each experiment is run for many trials, we end up with one value of $\alpha^*$ for each trial. Thus, we typically report the mode value of $\alpha^*$ across all trials.

\subsection{CBA Delegation Graphs}
\label{sec:CBA_experiments}

\begin{figure}[t]
    \centering
    \includegraphics[width=\columnwidth,keepaspectratio]{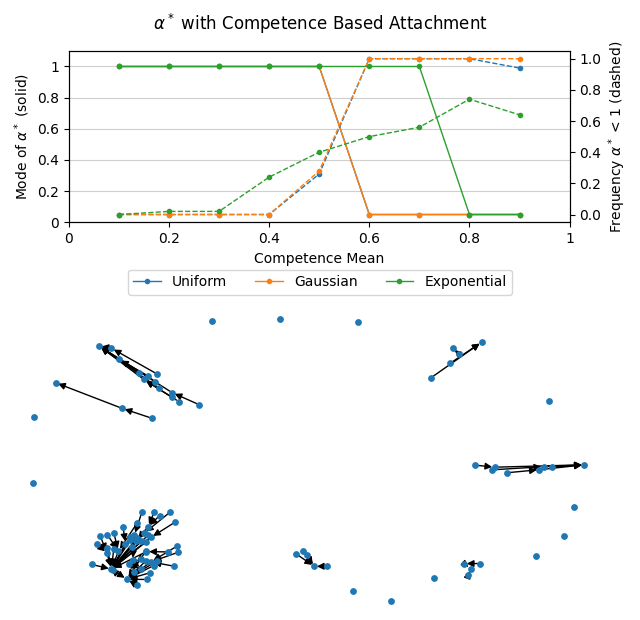}
    \caption{(Top) The mode value of $\alpha^*$ (solid) and frequency with which $\alpha^*$ is strictly below $1$ (dashed) over 50 trials on randomly generated CBA delegation graphs of 100 voters. A typical CBA graph (Bottom) is dominated by a few gurus receiving the large majority of delegations. As $\alpha$ increases the most powerful gurus become dictators and beneficial jury-theorem effects are lost.}
    \label{fig:cba_optimal_alpha_aggregate}
\end{figure}

\autoref{fig:cba_optimal_alpha_aggregate} shows how $\alpha^*$ changes as the average voter competence increases when voter delegations use a CBA topology. Each point represents 50 trials with 100 voters each. 

As CBA graphs are dominated by a small number of powerful gurus, \autoref{fig:cba_optimal_alpha_aggregate} shows a phase shift when mean competence reaches roughly $0.5$ At $\mu < 0.5$ most trials find $\alpha^* = 1$, corresponding to dominance from the large components in the graph. At $\mu > 0.5$, it becomes beneficial to have more gurus affecting the outcome (due to Jury Theorem effects; more voters is beneficial when average competence is above $0.5$) so $\alpha^*$ approaches 0.


\subsection{Viscosity in Random Networks}
\label{sec:optimal_alpha_delegating}
 
\begin{figure*}
    \centering
    \includegraphics[width=\textwidth,keepaspectratio]{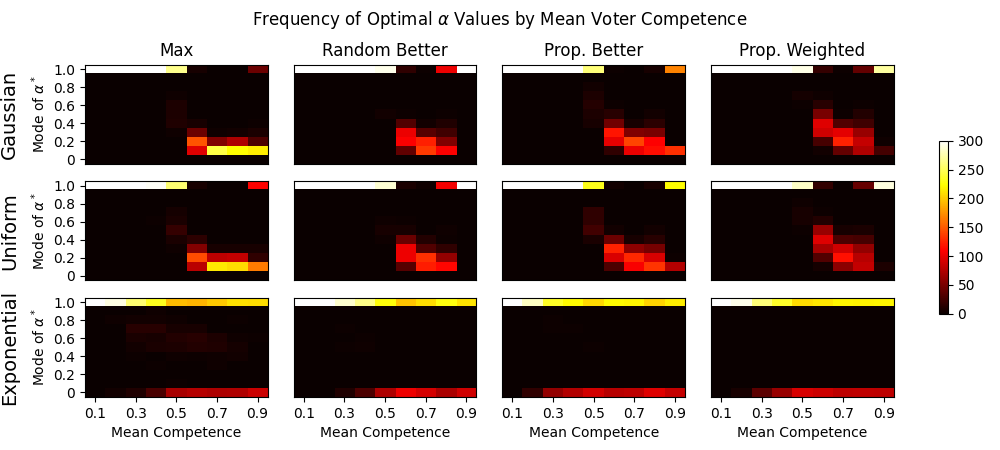}
    \caption{Distribution of $\alpha^*$ across competence distributions and delegation mechanisms. 300 trials were performed for each mean voter competence value. Each cell shows the number of trials at each mean competence value in which the corresponding $\alpha$ value is optimal. Results are on an a 100 voter Erdős–Rényi network with $p=0.1$ which is randomly regenerated at each trial. Similar results for Barabási–Albert networks are shown in \cref{appendix:experiment_results-heatmaps}.}
    \label{fig:opt_alpha_er_heatmap}
\end{figure*}

For each delegation mechanism and competence distribution, we ran 300 trials on randomly generated ER and BA social networks. \autoref{fig:opt_alpha_er_heatmap} shows how often each value of $\alpha$ is optimal for ER networks with $p=0.1$.

The heatmap immediately makes clear that $\alpha^*$ takes on a wide range of values. It is often $1$, but is often below $1$ as well. In particular, we see a shift similar to the CBA topology in \autoref{sec:CBA_experiments}: when mean competence is below 0.5, $\alpha^*$ is usually 1. When voters become more competent, $\alpha^*$ takes on a wider range of values between $0$ and $0.5$.

\begin{figure}
    \centering
    \includegraphics[width=\columnwidth,keepaspectratio]{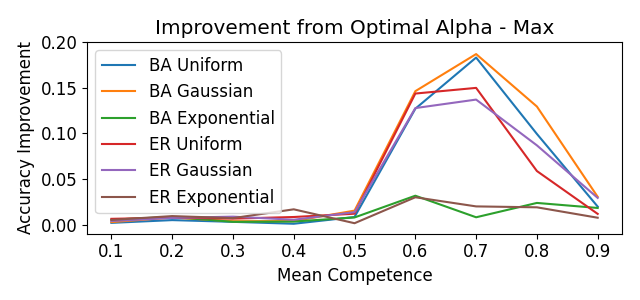}
    \caption{Accuracy improvement from using optimal viscosity vs no viscosity ($\alpha = 1$) for voters on random graph models delegating using the Max delegation mechanism. Results are averaged over 30 trials with 100 voters per experiment.}
    \label{fig:improvement_from_alpha}
\end{figure}

Moreover, \autoref{fig:improvement_from_alpha} shows the amount by which accuracy can be improved simply by setting $\alpha$ optimally. \autoref{fig:improvement_from_alpha} displays the difference between accuracy when $\alpha$ is set to 1 and when it is set to the approximate optimal value. When evaluating Uniform and Gaussian competence distributions, mean competence over 0.5 show significant benefit from optimal values of alpha while only mild improvement is seen with Exponential competencies. This effect holds across network type and delegation mechanism, as seen in \cref{appendix:experimental_results-accuracy_improvement}.

\subsection{Delegation Graph Structure}

We have examined several properties of the delegation graph structure to see whether they have an impact on the value of optimal alpha. From the delegation graphs of the same experiments discussed in \autoref{sec:optimal_alpha_delegating} we computed the values of several features. On this dataset, with $\alpha^*$ as a target variable, we ran a RandomForest regression. This found that no single feature was extremely predictive of $\alpha^*$ but mean competence was the \textit{most} predictive. Specifically, higher competence values were related with lower values of $\alpha^*$. A list of features we explored, and their importances is included in \cref{appendix:feature_importance}.

\subsection{Empirical Data}

To strengthen the results we show in \autoref{sec:optimal_alpha_delegating} we have run delegation experiments over several real-world networks. The networks are described in detail in~\cref{appendix:real_world_networks} and cover a wide range of sizes and topologies. These experiments show very similar distributions of $\alpha^*$, suggesting that these results are highly robust to network structure.

\section{Discussion}\label{section:discussion}

This paper shows that \textit{viscous democracy} significantly improves upon liquid democracy in many realistic scenarios. On simulated and empirical networks, when voters tend to have moderate competence, a low value for $\alpha$ is likely to improve the chance of retrieving the ground truth. Intuitively, viscosity increase the number of gurus that can affect the election by making it harder for voters to amass large amounts of power.
In contexts where there is a joint decision making objective and voter delegation, the use of viscosity should be strongly considered - it may provide up to 10\%-20\% improvement in accuracy over standard liquid democracy. 

\section{Outlook}
\label{section:outlook}

Our results show that viscosity improves decision-making ability which opens up several avenues for future research on why viscosity is useful and how to use this knowledge.

\begin{itemize}

\item
\textbf{Evaluating viscous democracy in other settings}:
Here we concentrated on a simple epistemic model of elections in which the decision to be made is binary. Evaluating the improvement from viscous democracy in more involved settings is important. Many settings remain open for exploration, such as multi-winner elections and more subjective settings where there is no ground truth.

\item
\textbf{Developing practical tools for viscous democracy}:
We have shown that viscosity typically improves delegation performance. Practically, developing tools that can efficiently predict the values of $\alpha^*$ across different settings is important. Various machine learning and optimization techniques may prove valuable for this task.

\end{itemize}

\bibliographystyle{named}
\bibliography{bibliography}

\begin{thebibliography}{}

\bibitem[\protect\citeauthoryear{Albert and Barab{\'a}si}{2002}]{albert2002statistical}
R{\'e}ka Albert and Albert-L{\'a}szl{\'o} Barab{\'a}si.
\newblock Statistical mechanics of complex networks.
\newblock {\em Reviews of modern physics}, 74(1):47, 2002.

\bibitem[\protect\citeauthoryear{Alouf-Heffetz \bgroup \em et al.\egroup }{2022}]{alouf2022should}
Shiri Alouf-Heffetz, Ben Armstrong, Kate Larson, and Nimrod Talmon.
\newblock How should we vote? a comparison of voting systems within social networks.
\newblock In {\em Proceedings of IJCAI '22}, pages 31--38, 2022.

\bibitem[\protect\citeauthoryear{Battiston \bgroup \em et al.\egroup }{2020}]{battiston2020networks}
Federico Battiston, Giulia Cencetti, Iacopo Iacopini, Vito Latora, Maxime Lucas, Alice Patania, Jean-Gabriel Young, and Giovanni Petri.
\newblock Networks beyond pairwise interactions: Structure and dynamics.
\newblock {\em Physics Reports}, 874:1--92, 2020.

\bibitem[\protect\citeauthoryear{Becker \bgroup \em et al.\egroup }{2021}]{becker2021can}
Ruben Becker, Gianlorenzo D'Angelo, Esmaeil Delfaraz, and Hugo Gilbert.
\newblock When can liquid democracy unveil the truth?
\newblock {\em CoRR}, abs/2104.01828, 2021.

\bibitem[\protect\citeauthoryear{Berkhin}{2005}]{berkhin2005survey}
Pavel Berkhin.
\newblock A survey on pagerank computing.
\newblock {\em Internet mathematics}, 2(1):73--120, 2005.

\bibitem[\protect\citeauthoryear{Bloembergen \bgroup \em et al.\egroup }{2019}]{bloembergen2019rational}
Daan Bloembergen, Davide Grossi, and Martin Lackner.
\newblock On rational delegations in liquid democracy.
\newblock In {\em Proceedings of AAAI '19'}, volume~33, pages 1796--1803, 2019.

\bibitem[\protect\citeauthoryear{Blum and Zuber}{2016}]{blum2016liquid}
Christian Blum and Christina~Isabel Zuber.
\newblock Liquid democracy: Potentials, problems, and perspectives.
\newblock {\em Journal of Political Philosophy}, 24(2):162--182, 2016.

\bibitem[\protect\citeauthoryear{Boldi \bgroup \em et al.\egroup }{2011}]{boldi2011viscous}
Paolo Boldi, Francesco Bonchi, Carlos Castillo, and Sebastiano Vigna.
\newblock Viscous democracy for social networks.
\newblock {\em Communications of the ACM}, 54(6):129--137, 2011.

\bibitem[\protect\citeauthoryear{Boldi \bgroup \em et al.\egroup }{2015}]{boldi2015liquid}
Paolo Boldi, Corrado Monti, Massimo Santini, and Sebastiano Vigna.
\newblock Liquid fm: recommending music through viscous democracy.
\newblock {\em arXiv preprint arXiv:1503.08604}, 2015.

\bibitem[\protect\citeauthoryear{Brill and Talmon}{2018}]{brill2018pairwise}
Markus Brill and Nimrod Talmon.
\newblock Pairwise liquid democracy.
\newblock In {\em Proceedings of IJCAI '18}, volume~18, pages 137--143, 2018.

\bibitem[\protect\citeauthoryear{Burkardt}{2014}]{burkardt2014truncated}
John Burkardt.
\newblock The truncated normal distribution.
\newblock {\em Department of Scientific Computing Website, Florida State University}, pages 1--35, 2014.

\bibitem[\protect\citeauthoryear{Caragiannis and Micha}{2019}]{caragiannis2019contribution}
Ioannis Caragiannis and Evi Micha.
\newblock A contribution to the critique of liquid democracy.
\newblock In {\em IJCAI}, pages 116--122, 2019.

\bibitem[\protect\citeauthoryear{Dey \bgroup \em et al.\egroup }{2021}]{dey2021parameterized}
Palash Dey, Arnab Maiti, and Amatya Sharma.
\newblock On parameterized complexity of liquid democracy.
\newblock In {\em Proceedings of CALDAM '21}, pages 83--94, 2021.

\bibitem[\protect\citeauthoryear{Erdos \bgroup \em et al.\egroup }{1960}]{erdos1960evolution}
Paul Erdos, Alfr{\'e}d R{\'e}nyi, et~al.
\newblock On the evolution of random graphs.
\newblock {\em Publ. Math. Inst. Hung. Acad. Sci}, 5(1):17--60, 1960.

\bibitem[\protect\citeauthoryear{Escoffier \bgroup \em et al.\egroup }{2019}]{escoffier2019convergence}
Bruno Escoffier, Hugo Gilbert, and Ad{\`e}le Pass-Lanneau.
\newblock The convergence of iterative delegations in liquid democracy in a social network.
\newblock In {\em Proceedings of SAGT '19}, pages 284--297, 2019.

\bibitem[\protect\citeauthoryear{Gleiser and Danon}{2003}]{gleiser2003community}
Pablo~M Gleiser and Leon Danon.
\newblock Community structure in jazz.
\newblock {\em Advances in complex systems}, 6(04):565--573, 2003.

\bibitem[\protect\citeauthoryear{Gonzalez}{1985}]{GONZALEZ1985293}
Teofilo~F. Gonzalez.
\newblock Clustering to minimize the maximum intercluster distance.
\newblock {\em Theoretical Computer Science}, 38:293--306, 1985.

\bibitem[\protect\citeauthoryear{Green-Armytage}{2015}]{green2015direct}
James Green-Armytage.
\newblock Direct voting and proxy voting.
\newblock {\em Constitutional Political Economy}, 26:190--220, 2015.

\bibitem[\protect\citeauthoryear{Grofman \bgroup \em et al.\egroup }{1983}]{grofman1983thirteen}
Bernard Grofman, Guillermo Owen, and Scott~L Feld.
\newblock Thirteen theorems in search of the truth.
\newblock {\em Theory and decision}, 15(3):261--278, 1983.

\bibitem[\protect\citeauthoryear{Guimera \bgroup \em et al.\egroup }{2003}]{guimera2003self}
Roger Guimera, Leon Danon, Albert Diaz-Guilera, Francesc Giralt, and Alex Arenas.
\newblock Self-similar community structure in a network of human interactions.
\newblock {\em Physical review E}, 68(6):065103, 2003.

\bibitem[\protect\citeauthoryear{Halpern \bgroup \em et al.\egroup }{2021}]{halpern2021defense}
Daniel Halpern, Joseph~Y Halpern, Ali Jadbabaie, Elchanan Mossel, Ariel~D Procaccia, and Manon Revel.
\newblock In defense of liquid democracy.
\newblock {\em arXiv preprint arXiv:2107.11868}, 2021.

\bibitem[\protect\citeauthoryear{Hassan and De~Filippi}{2021}]{hassan2021decentralized}
Samer Hassan and Primavera De~Filippi.
\newblock Decentralized autonomous organization.
\newblock {\em Internet Policy Review}, 10(2):1--10, 2021.

\bibitem[\protect\citeauthoryear{Jain \bgroup \em et al.\egroup }{2022}]{jain2022preserving}
Pallavi Jain, Krzysztof Sornat, and Nimrod Talmon.
\newblock Preserving consistency for liquid knapsack voting.
\newblock In {\em Proceedings of ECAI '22}, pages 221--238, 2022.

\bibitem[\protect\citeauthoryear{Kahng \bgroup \em et al.\egroup }{2021}]{kahng2021liquid}
Anson Kahng, Simon Mackenzie, and Ariel Procaccia.
\newblock Liquid democracy: An algorithmic perspective.
\newblock {\em Journal of Artificial Intelligence Research}, 70:1223--1252, 2021.

\bibitem[\protect\citeauthoryear{Kleinberg}{2010}]{Kleinberg-networks-book}
Jon Kleinberg.
\newblock Ect volume 26 issue 5 cover and back matter.
\newblock {\em Econometric Theory}, 26(5):b1–b4, 2010.

\bibitem[\protect\citeauthoryear{Knuth}{1993}]{knuth1993stanford}
Donald~Ervin Knuth.
\newblock {\em The Stanford GraphBase: a platform for combinatorial computing}, volume~1.
\newblock AcM Press New York, 1993.

\bibitem[\protect\citeauthoryear{Lusseau \bgroup \em et al.\egroup }{2003}]{lusseau2003bottlenose}
David Lusseau, Karsten Schneider, Oliver~J Boisseau, Patti Haase, Elisabeth Slooten, and Steve~M Dawson.
\newblock The bottlenose dolphin community of doubtful sound features a large proportion of long-lasting associations: can geographic isolation explain this unique trait?
\newblock {\em Behavioral Ecology and Sociobiology}, 54:396--405, 2003.

\bibitem[\protect\citeauthoryear{May}{1952}]{may1952set}
Kenneth~O May.
\newblock A set of independent necessary and sufficient conditions for simple majority decision.
\newblock {\em Econometrica: Journal of the Econometric Society}, pages 680--684, 1952.

\bibitem[\protect\citeauthoryear{Newman}{2006}]{newman2006finding}
Mark~EJ Newman.
\newblock Finding community structure in networks using the eigenvectors of matrices.
\newblock {\em Physical review E}, 74(3):036104, 2006.

\bibitem[\protect\citeauthoryear{Paulin}{2020}]{paulin2020overview}
Alois Paulin.
\newblock An overview of ten years of liquid democracy research.
\newblock In {\em DG.O '20}, pages 116--121, 2020.

\bibitem[\protect\citeauthoryear{Pedregosa \bgroup \em et al.\egroup }{2011}]{scikit-learn}
F.~Pedregosa, G.~Varoquaux, A.~Gramfort, V.~Michel, B.~Thirion, O.~Grisel, M.~Blondel, P.~Prettenhofer, R.~Weiss, V.~Dubourg, J.~Vanderplas, A.~Passos, D.~Cournapeau, M.~Brucher, M.~Perrot, and E.~Duchesnay.
\newblock Scikit-learn: Machine learning in {P}ython.
\newblock {\em Journal of Machine Learning Research}, 12:2825--2830, 2011.

\bibitem[\protect\citeauthoryear{Shapiro}{2018}]{shapiro2018point}
Ehud Shapiro.
\newblock Point: foundations of e-democracy.
\newblock {\em Communications of the ACM}, 61(8):31--34, 2018.

\bibitem[\protect\citeauthoryear{Watts and Strogatz}{1998}]{watts1998collective}
Duncan~J Watts and Steven~H Strogatz.
\newblock Collective dynamics of ‘small-world’networks.
\newblock {\em nature}, 393(6684):440--442, 1998.

\bibitem[\protect\citeauthoryear{Zachary}{1977}]{zachary1977information}
Wayne~W Zachary.
\newblock An information flow model for conflict and fission in small groups.
\newblock {\em Journal of anthropological research}, 33(4):452--473, 1977.

\bibitem[\protect\citeauthoryear{Zhang and Grossi}{2021}]{zhang2021power}
Yuzhe Zhang and Davide Grossi.
\newblock Power in liquid democracy.
\newblock In {\em Proceedings AAAI '21}, volume~35, pages 5822--5830, 2021.

\end{thebibliography}

\newpage
\clearpage

\appendix

\section{Delegation Functions}
\label{appendix:delegation_function_descriptions}

\begin{itemize}

\item
\textbf{Max}:
Each delegating voter $v_i$ delegates to their most competent neighbour.

\[
p^\text{max}(v_i, v_j) = \begin{cases} 
      1 & q_j = \argmax_j \in \{q_j \in Q | (i, j) \in E\} \\
      0 & \text{otherwise}
   \end{cases}
\]

\item
\textbf{Random Better}:
Each delegating voter, $v_i$, selects one of their neighbours uniformly at random from the set of neighbours with higher competence. That is, $v_i$ delegates to a random voter from the set $N^{+}_G\brackets{i}$ where:
\[
N^{+}_G(i) = \left\{j\in V \mid (i,j)\in E\;, \; q_j > q_i \right\}
\]
\[
p^\text{rand\_better}(v_i, v_j) = \begin{cases} 
      \frac{1}{|N^+(v_i)|} & j \in N^+(v_i) \\
      0 & \text{otherwise}
   \end{cases}
\]

\item
\textbf{Proportional Better}:
Delegators delegate to a neighbour with higher competence, however, the chance of delegating to a neighbour is directly correlated with the difference between their guru's competence and that of the delegator.

\[
p^\text{prop\_better}(v_i, v_j) \propto \begin{cases} 
      \frac{q_{d^*(j)} - q_i}{\sum_{v_k \in N^+(v_i)} q_{d^*(k)}-q_i } & j \in N^+(v_i) \\
      0 & \text{otherwise}
   \end{cases}
\]

\item
\textbf{Proportional Weighted}:
Delegation probabilities are based on both the competence difference between delegator and delegatee, as well as the weight of the representative ultimately being delegated to. A lower weight leads to a higher delegation probability.

\[
p^\text{prop\_weighted}(v_i, v_j) \propto \begin{cases} 
      \clubsuit & j \in N^+(v_i) \\
      0 & \text{otherwise}
   \end{cases}
\]
where $\clubsuit := \frac{1}{w_{d^*(j)}} \frac{q_{d^*(j)} - q_i}{\sum_{v_k \in N^+(v_i)} q_{d^*(k)}-q_i }$.

\end{itemize}

\section{Estimating $\alpha^*$}
\label{appendix:estimating_optimal_alpha}


\begin{algorithm}[h!]
\caption{Estimating $\alpha^*$}
\label{alg:algorithm_opt_alpha}

\raggedright
\textbf{Input}: $t, k, Q, D$

\begin{algorithmic}[1] 
\STATE $\alpha^*, \mu_{\text{max}}, \sigma_{\text{max}} \gets 1, 1, 1$
\STATE $A \gets t$ points sampled evenly from $[0, 1]$
\FOR{$a_t \in A$}
    \STATE $\alpha \gets a_t$
    \FOR{$i \in 0..k$}
    \STATE $acc_{t,i} \gets Acc(\alpha, Q, D)$
    \ENDFOR
    \STATE $\mu_t, \sigma_t \gets \text{mean}(acc_{t, -}), \text{st. dev.}(acc_{t, -})$
    \IF{$\mu_t - \sigma_t > \mu_{\text{max}} + \sigma_{\text{max}}$}
    \STATE $\alpha^*, \mu_{\text{max}}, \sigma_{\text{max}} \gets \alpha, \mu_t, \sigma_t$
    \ENDIF
\ENDFOR
    
\end{algorithmic}
\end{algorithm}

\newpage
\section{Description of Real-World Networks}
\label{appendix:real_world_networks}

\begin{table}[h!]
\centering
\begin{tabular}{@{}llll@{}}
\toprule
\multicolumn{1}{c}{Network} & \multicolumn{1}{c}{Nodes} & \multicolumn{1}{c}{Edges} & \multicolumn{1}{c}{Source}     \\ \midrule
celegans                   & 296                       & 2359                      & \cite{watts1998collective}    \\
dolphins                         & 62                        & 159                       & \cite{lusseau2003bottlenose}  \\
email                            & 1133                      & 10903                     & \cite{guimera2003self}        \\
jazz                             & 198                       & 5484                      & \cite{gleiser2003community}   \\
netscience                       & 1589                      & 2700                      & \cite{newman2006finding}      \\
karate                           & 34                        & 78                        & \cite{zachary1977information} \\
lesmis                           & 77                        & 254                       & \cite{knuth1993stanford}      \\ \bottomrule
\end{tabular}
\caption{The real-world networks used in our analysis.}
\label{figure:realworldnetworks}
\end{table}


\section{Supplemental Experiment Results}
\label{appendix:experiment_results}

\subsection{Distribution of Optimal $\alpha^*$}
\label{appendix:experiment_results-heatmaps}

\begin{figure*}[h!]
    \centering
    \includegraphics[width=\textwidth,keepaspectratio]{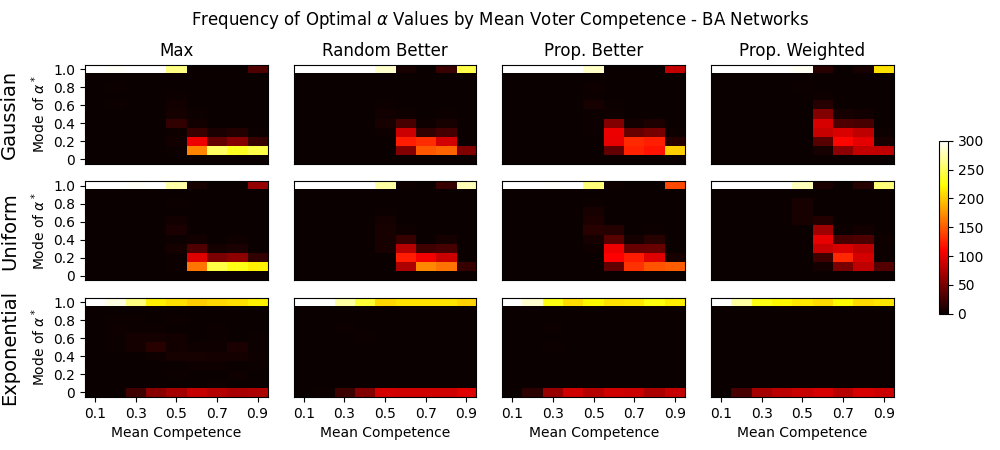}
    \caption{Distribution of $\alpha^*$ across competence distributions and delegation mechanisms. 300 trials were performed for each mean voter competence value. Each cell shows the fraction of trials at each mean competence value in which the corresponding $\alpha$ value is optimal. Results are on a 100 voter Barabási–Albert network with $m=10$ which is randomly regenerated at each trial.}
    \label{fig:opt_alpha_ba_heatmap}
\end{figure*}

\begin{figure*}[h]
    \centering
    \includegraphics[width=\textwidth,keepaspectratio]{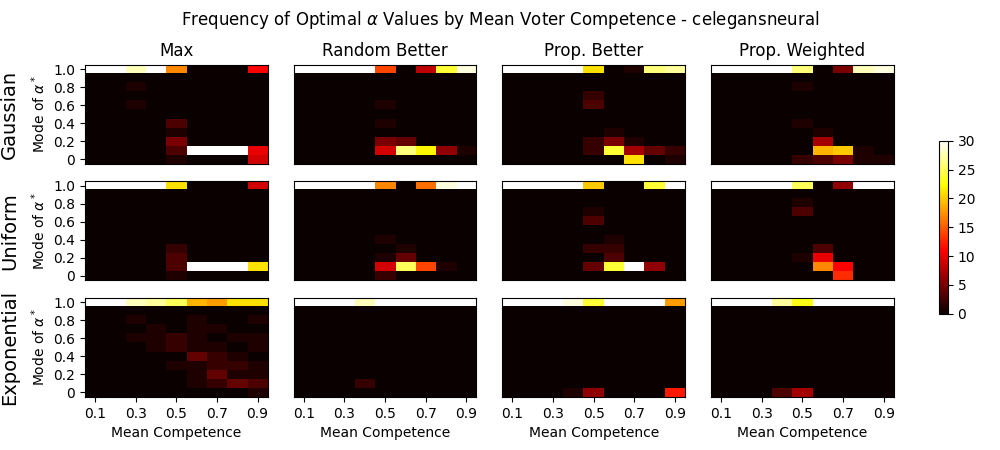}
    \caption{Distribution of $\alpha^*$ across competence distributions and delegation mechanisms. 30 trials were performed for each mean voter competence value. Each cell shows the fraction of trials at each mean competence value in which the corresponding $\alpha$ value is optimal. Results are shown for the \textbf{celegansneural} network.}
    \label{fig:opt_alpha_celegansneural_heatmap}
\end{figure*}

\begin{figure*}[h]
    \centering
    \includegraphics[width=\textwidth,keepaspectratio]{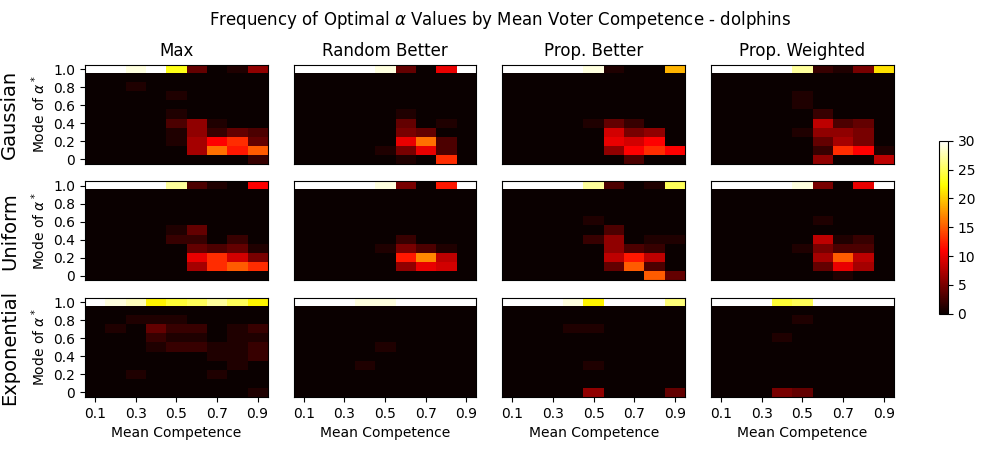}
    \caption{Distribution of $\alpha^*$ across competence distributions and delegation mechanisms. 30 trials were performed for each mean voter competence value. Each cell shows the fraction of trials at each mean competence value in which the corresponding $\alpha$ value is optimal. Results are shown for the \textbf{dolphins} network.}
    \label{fig:opt_alpha_dolphins_heatmap}
\end{figure*}

\begin{figure*}[h]
    \centering
    \includegraphics[width=\textwidth,keepaspectratio]{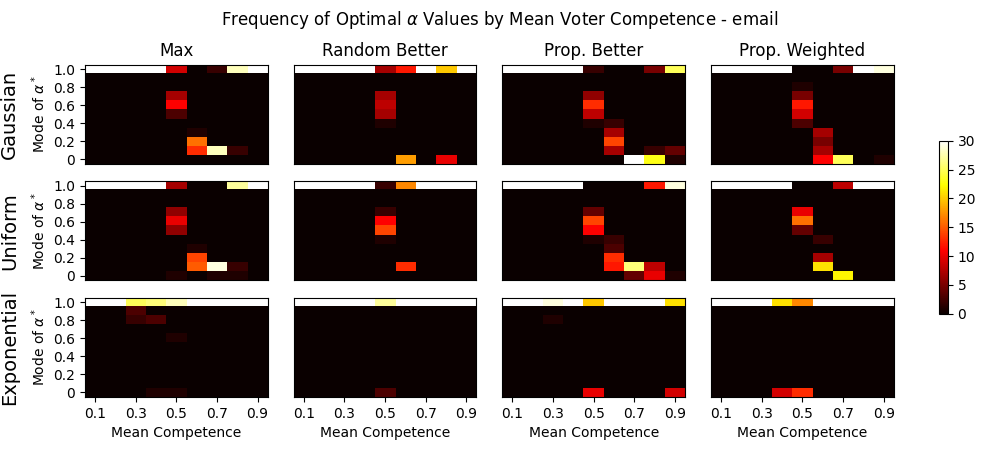}
    \caption{Distribution of $\alpha^*$ across competence distributions and delegation mechanisms. 30 trials were performed for each mean voter competence value. Each cell shows the fraction of trials at each mean competence value in which the corresponding $\alpha$ value is optimal. Results are shown for the \textbf{email} network.}
    \label{fig:opt_alpha_email_heatmap}
\end{figure*}

\begin{figure*}[h]
    \centering
    \includegraphics[width=\textwidth,keepaspectratio]{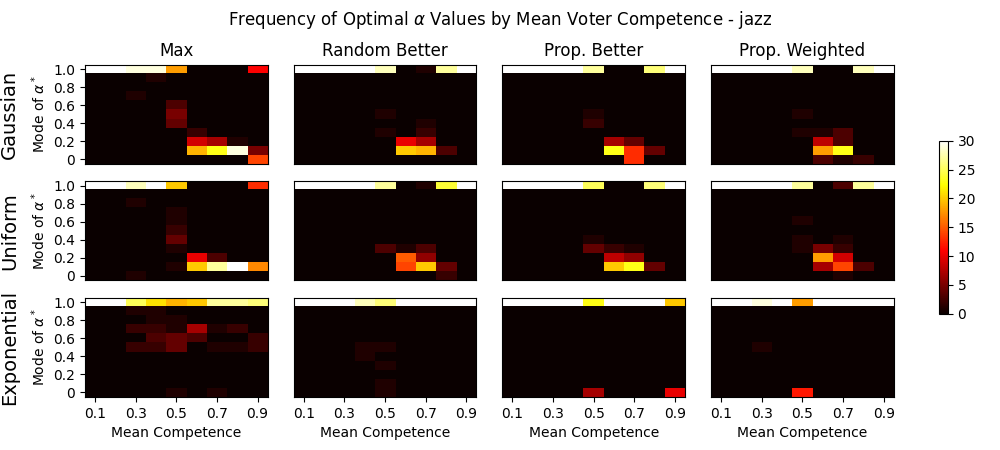}
    \caption{Distribution of $\alpha^*$ across competence distributions and delegation mechanisms. 30 trials were performed for each mean voter competence value. Each cell shows the fraction of trials at each mean competence value in which the corresponding $\alpha$ value is optimal. Results are shown for the \textbf{jazz} network.}
    \label{fig:opt_alpha_jazz_heatmap}
\end{figure*}

\begin{figure*}[h]
    \centering
    \includegraphics[width=\textwidth,keepaspectratio]{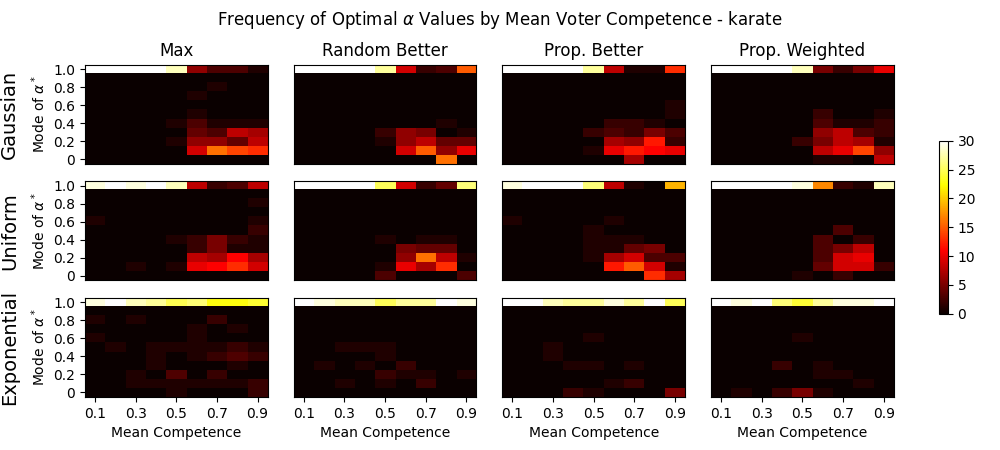}
    \caption{Distribution of $\alpha^*$ across competence distributions and delegation mechanisms. 30 trials were performed for each mean voter competence value. Each cell shows the fraction of trials at each mean competence value in which the corresponding $\alpha$ value is optimal. Results are shown for the \textbf{karate} network.}
    \label{fig:opt_alpha_karate_heatmap}
\end{figure*}

\begin{figure*}[h]
    \centering
    \includegraphics[width=\textwidth,keepaspectratio]{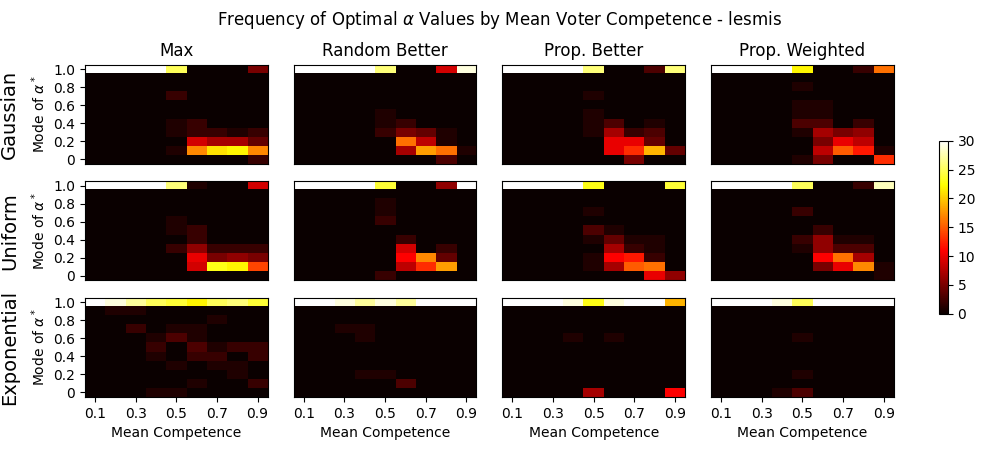}
    \caption{Distribution of $\alpha^*$ across competence distributions and delegation mechanisms. 30 trials were performed for each mean voter competence value. Each cell shows the fraction of trials at each mean competence value in which the corresponding $\alpha$ value is optimal. Results are shown for the \textbf{lesmis} network.}
    \label{fig:opt_alpha_lesmis_heatmap}
\end{figure*}

\begin{figure*}[h]
    \centering
    \includegraphics[width=\textwidth,keepaspectratio]{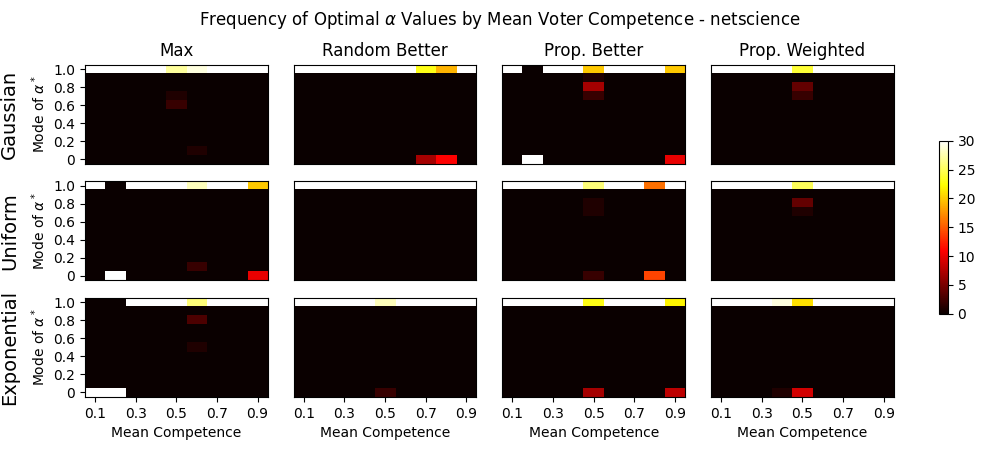}
    \caption{Distribution of $\alpha^*$ across competence distributions and delegation mechanisms. 30 trials were performed for each mean voter competence value. Each cell shows the fraction of trials at each mean competence value in which the corresponding $\alpha$ value is optimal. Results are shown for the \textbf{netscience} network.}
    \label{fig:opt_alpha_netscience_heatmap}
\end{figure*}

\clearpage
\newpage
\subsection{Accuracy Improvement from $\alpha^*$}
\label{appendix:experimental_results-accuracy_improvement}

\begin{figure}[h!]
    \centering
    \includegraphics[width=\columnwidth,keepaspectratio]{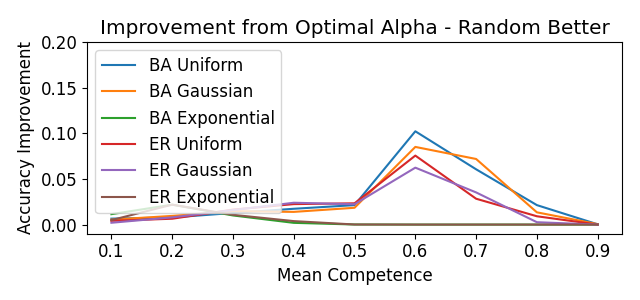}
    \caption{Accuracy improvement from using optimal viscosity vs no viscosity ($\alpha = 1$) for voters on random graph models delegating using the \textbf{Random Better} mechanism. Results are averaged over 30 trials with 100 voters per experiment.}
    \label{fig:improvement_from_alpha-random_better}
\end{figure}

\begin{figure}[h!]
    \centering
    \includegraphics[width=\columnwidth,keepaspectratio]{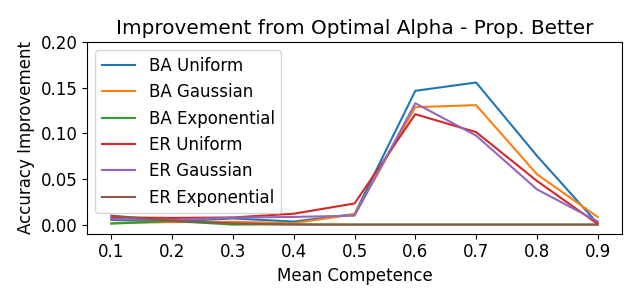}
    \caption{Accuracy improvement from using optimal viscosity vs no viscosity ($\alpha = 1$) for voters on random graph models delegating using the \textbf{Proportional Better} mechanism. Results are averaged over 30 trials with 100 voters per experiment.}
    \label{fig:improvement_from_alpha-proportional_better}
\end{figure}

\begin{figure}[h!]
    \centering
    \includegraphics[width=\columnwidth,keepaspectratio]{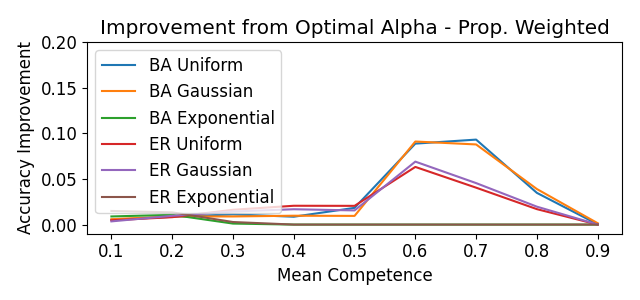}
    \caption{Accuracy improvement from using optimal viscosity vs no viscosity ($\alpha = 1$) for voters on random graph models delegating using the \textbf{Proportional Weighted} mechanism. Results are averaged over 30 trials with 100 voters per experiment.}
    \label{fig:improvement_from_alpha-proportional_weighted}
\end{figure}

\newpage
\subsection{Feature Importance}
\label{appendix:feature_importance}

\begin{figure}[h]
    \centering
    \includegraphics[width=0.5\textwidth,keepaspectratio]{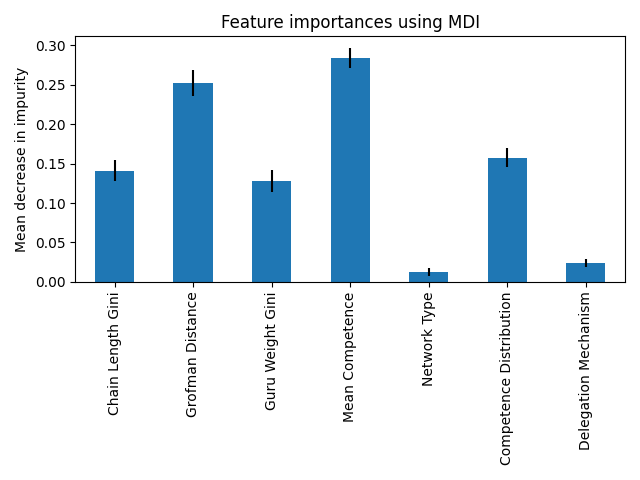}
    \caption{Here we plotted the feature importance given a classification of $\alpha$ using random forest from the previous experiments in \autoref{sec:experiments}.}
    \label{fig:feature importance}
\end{figure}

We used a Random Forest regression model to explore whether any features of an election were predictive of $\alpha^*$. We assembled values of the features listed below along with their corresponding $\alpha^*$. These values were used to train a sci-kit learn RandomForestRegressor model \cite{scikit-learn} which reached an $R^2$ value of approximately 0.6, indicating the features had some predictive power but very little. We then computed the relative importance of each feature in the model, shown in \autoref{fig:feature importance}. Mean voter competence is most predictive of $\alpha^*$ and has an inverse relationship: Higher mean competence loosely indicates a lower value of $\alpha^*$.

The features used in this regression are described below:

\begin{itemize}
    \item Chain Length Gini - For each guru, measure the distance from the guru to each of its delegators. Take the Gini index of all these lengths.
    \item Grofman Distance - The $\ell2$ distance of the current weight distribution of the gurus compared to the ideal $w_i\sim log\brackets{\frac{q_i}{1-q_i}}$ (see \cite{grofman1983thirteen}).
    \item Guru Weight Gini - The Gini index of the guru weights.
    \item Mean Competence - The mean of the underlying competence distribution.
    \item Network Type - Possible values: BA and ER.
    \item Competence Distribution - Possible values: Uniform, Gaussian or Exponential.
    \item Delegation Mechanism - Possible values: Max, Proportional Better, Random Better, Proportional Weighted.
\end{itemize}

\end{document}